\documentstyle[12pt,epsfig]{article}
\begin{document}
\begin{center}

{\large{\bf Light Meson Decay Constants}}

\vspace*{5mm}

\underline{G.~Ganbold}\footnote{E--mail:{\tt ~~ganbold@thsun1.jinr.ru}}

\vspace*{3mm}

{\sl Bogoliubov Lab. Theor. Phys., JINR, 141980, Dubna, Russia} \\
{\sl Institute of Physics and Technology, 210651, Ulaanbaatar, Mongolia}
\end{center}

\begin{abstract}
We estimate the decay constants $f_\pi$ and $f_K$ of the light
mesons within a relativistic quantum-field model of interacting
quarks and gluons confined analytically. The necessary physical
parameters, the quark masses $m_u$ and $m_s$, the coupling constant
$\alpha_s$ and the confinement scale $\Lambda$ have been obtained
from our previous investigation on the meson ground states, orbital
and radial excitations by using the ladder Bethe-Salpeter equation.
Our model provides a solid framework to compute the meson spectra,
the lowest glueball state and decay constants of light mesons from
the basic principles of QCD and QFT.
\end{abstract}

\vskip 3mm



\vskip 3mm

\section{Introduction}

The pseudoscalar mesons, especially the $\pi$ and $K$, are studied
extensively to understand the under-structures from nonperturbative
QCD (e.g., \cite{robe03}). The decay constants of the light mesons
are one  of the basic parameters in the particle physics and form an
extremely important test ground for any nonperturbative QCD methods.
Particularly, the pseudoscalar-meson decay constants $f_P$ play an
important role in hadron phenomenology (e.g., the leptonic decay width
is proportional to $f_P^2$ and the $V_{us}$ element is extracted from
the ratio $f_K/f_\pi$) and can be used as a test ground for models.

A number of theoretical approaches obeying an accuracy comparable
to the uncertainties in the experimental data, has relied on models
containing too many, in some cases even nonphysical, parameters.
On the other hand, the scaling property of QCD implies that all
characteristics of hadrons must depend on a few global parameters.

Among the significant theoretical approaches, the Bethe-Salpeter
equation (BSE) is a conventional approach in dealing with the two-body
relativistic bound state problems \cite{salp51} and its solutions
give useful information about the under-structure of the mesons. Many
analytical and numerical calculations indicate that the ladder BSE with
phenomenological models can give model independent results successful
descriptions of the long distance properties of the low energy QCD and
the QCD vacuum  \cite{alko01,efim02,robe03}.

Below we adopt a relativistic quantum field model of interacting
quarks and gluons under the analytic confinement, solve the BSE in
the one-gluon exchange approximation for the quark-antiquark bound
states and estimate the decay constants $f_\pi$ and $f_K$. The model
relies on only necessary physical parameters $\{m_f,\alpha_s,\Lambda\}$
which have been fixed to fit the meson masses in the ground states,
orbital and radial excitations as well as the lowest-state glueball
mass \cite{ganb05,ganb06}.

\section{Two-quark Bound States}

In the conventional quark model mesons are $(q\bar{q'})$ bound states
of quarks and antiquarks (the flavors may be different). We deal with
the pseudoscalars $P$ ($J^{PC}=0^{-+}$) and vectors $V$ ($1^{--}$),
the most established sectors of hadron spectroscopy.

Consider a Yukawa-type interaction of quarks and gluons with the Lagrangian
\begin{eqnarray}
{\cal L}=
\left( {\bar\Psi}S^{-1}{\Psi}\right) + {1\over 2}\left( \phi~D^{-1}
\phi \right)+g\left({\bar\Psi_\alpha^i}
i\gamma_\mu^{\alpha\beta} t^a_{ij}{\Psi_\beta^j}\phi_\mu^a \right) \,,
\end{eqnarray}
where $\Psi_{\alpha f}^a(x)$ is the quark field, $\phi_\mu^a(x)$ is the
gluon field, $g$ is the coupling strength, $t^a$ are the Gell-Mann
matrices and $\{a,\alpha,f\}$ are the color, spin and flavor indices.

The conventional QCD encounters a difficulty by defining the explicit
quark and gluon propagator at the confinement scale \cite{mari03}.
Particularly, the infrared form of the gluon propagator is still a
controversial aspect \cite{robe94}, there are various versions,
with different behaviors for the infrared region: finite, zero and
more divergent that $1/k^2$, each one with its advantages and
inconveniences \cite{habe90,corn91,smek97,bloc01}.

Nevertheless, the matrix elements of hadron processes at large
distance are in fact integrated characteristics of the quark and
gluon functions and taking into account the correct global symmetry
properties and their breaking by introducing additional physical
parameters may be more important than the working out in detail
(e.g., \cite{feld00,efim02,ganb04}). Besides, the bound states of
quarks and gluons may be found as the solution of the BSE in a
variational-integral form \cite{ganb05,ganb06} that is low sensible
on tiny details of integrands.

We consider the simplest effective quark and gluon propagators:
\begin{eqnarray}
&& \tilde{S}_{\alpha\beta}^{ij}(\hat{p})=
~\delta^{ij} {\left\{ i\hat{p}+m~[1+\gamma_5\omega(m)]
\right\}_{\alpha\beta}\over m^2}
\exp\left\{-{p^2+m^2\over 2\Lambda^2} \right\}\,, \nonumber\\
&& D_{ab}^{\mu\nu}(x)=\delta_{ab}\delta^{\mu\nu}~{\Lambda^2\over (4\pi)^2}
~\exp\left\{-{x^2\Lambda^2\over 4}\right\}
= \delta_{ab}\delta^{\mu\nu}~D(x)  \,,
\label{propagat}
\end{eqnarray}
where $\hat{p}=p_\mu \gamma_\mu$, $m$ - the quark mass and
$0<\omega(m)<1$ \cite{ganb06}. These entire analytic functions in
the Euclidean metric keep the spirit of the {\sl analytic confinement}
\cite{leut81,efned95,efim02} and imply that each isolated quark and
gluon is confined in the background gluon field.

The meson mass is derived from \cite{ganb06}:
\begin{eqnarray}
1+\lambda_{{\cal N}}(M_{{\cal N}}^2)=0\,.
\label{Bethe2}
\end{eqnarray}
where the two-quark bound states are obtained by diagonalizing the
quadratic part in the meson partition function on the full orthonormal
system $\{U_{\cal N}\}$. This is nothing else but the
solution of the appropriate BSE (in the one-gluon approximation):
\begin{eqnarray}
\int\!\!\!\!\int\! dx dy ~U_{\cal N}(x)\left\{ 1+g^2\sqrt{D(x)}
~\Pi_{JJ'}(p^2,x,y)\sqrt{D(y)} \right\}
U_{\cal N'}(y)=\delta^{{\cal NN'}} [1+\lambda_{{\cal N}}(-p^2)] \,,
\label{Bethe1}
\end{eqnarray}
where, the two-point function is introduced:
\begin{eqnarray*}
\Pi_{JJ'}(p^2,x,y)= \int\! {d^4 k\over(2\pi)^4}~e^{-ik(x-y)}
{\bf Tr}\left[\Gamma_J \tilde{S}\left(\hat{k}+\mu_1\hat{p}\right)
\Gamma_{J'} \tilde{S}\left(\hat{k}-\mu_2\hat{p}\right) \right]
\label{twopoint}
\end{eqnarray*}
and $\Gamma_J=\{i\gamma_5,~i\gamma_\mu \}$ for $J=\{P,V\}$.

We have solved Eq.(\ref{Bethe2}) within a variational-integral approach
and have found that the following values of free parameters:
\begin{eqnarray}
\alpha_s=0.186 \,,\qquad  \Lambda=730 {\mbox{\rm ~MeV}} \,,
\qquad m_{\{u,~d\}}=170{\mbox{\rm ~MeV}} \,, \nonumber\\
m_s=188{\mbox{\rm ~MeV}}  \,,\qquad m_c=646{\mbox{\rm ~MeV}}  \,,
\qquad m_b=4221{\mbox{\rm ~MeV}} \,.
\label{parameters}
\end{eqnarray}
are optimal for the meson ground states, orbital and radial excitations
\cite{ganb06}. Our estimates for the $P$ and $V$ meson masses in the
ground state (Fig.1) were in good agreement with the experimental data
\cite{PDG2006} and have a small experimental error ($~1\div 5$ per cent)
in a wide range $\sim 140\div 9500{\mbox{\rm ~MeV}}$.

\begin{figure}[ht]
 \centerline{
 \includegraphics[width=100mm,height=80mm]{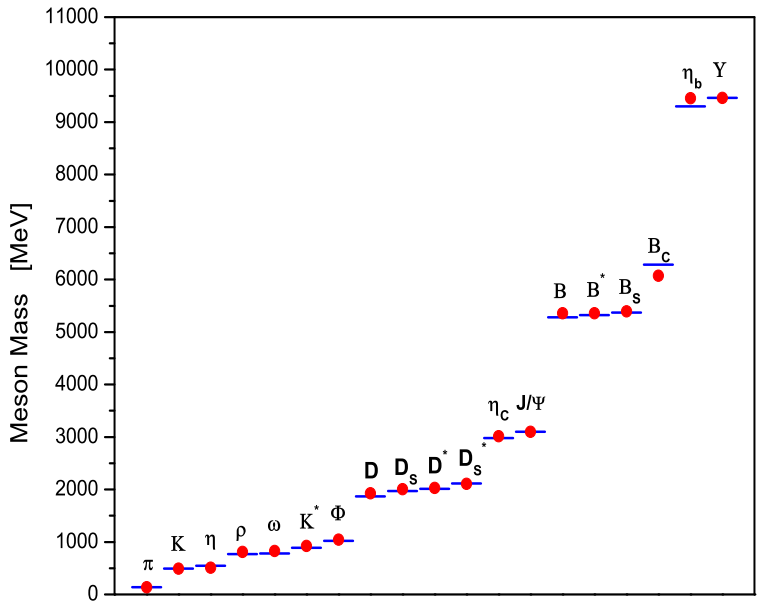} }
 \caption{Estimated meson masses (dots) compared with experimental data (dashes).}
\end{figure}

The lowest-state (scalar) glueball mass has been derived analytically
\cite{ganb06}:
\begin{eqnarray}
\label{glueball}
M_G^2 = 2\Lambda^2\ln\left({\alpha_{upp}\over\alpha_s}\right)=
\left( 1745{\mbox{\rm ~MeV}}\right)^2  \,,
\qquad \alpha_{upp}={2\pi(2+\sqrt{3})^2\over 27}\,.
\end{eqnarray}

Our estimate (\ref{glueball}) is close to the result
$M_G=1710\pm 100 {\mbox{\rm ~MeV}}$ due to the quantized knot soliton model
\cite{kond06} and quenched lattice estimate \cite{chen06} and in agreement
with the predictions expecting non-$q\bar{q}$ scalar object in the range
$\sim 1600\div 1800{\mbox{\rm ~MeV}}$ \cite{amsl04,bugg04}.

The estimated value $\alpha_s=0.186$ coincides with the latest experimental
data $\alpha_{s,(expr)}\approx 0.10\div 0.35$ \cite{PDG2006} and the
prediction of the quenched theory $\alpha_{s,(quench)}\approx 0.195$ \cite{kasz05}.
Note, this relatively weak coupling justifies the use of the one-gluon
exchange mode in our consideration. By analysing the glueball mass formula
(\ref{glueball}) one obtains the restriction $\alpha_s < \alpha_{upp}\approx 3.24124$
that coincides with the two-loop value of the freezing coupling constant
\cite{bada02}.

A rough estimate of the hadronization distance
$$
r_{hadr}^2 \sim {\int d^4 x~x^2~D(x) \over\int d^4
x~x~D(x)}={8\over\Lambda^2} \sim \left({1\over 250 \mbox{\rm MeV}}\right)^2
$$
shows that the confinement radius $r_{conf}\simeq 1/\Lambda$ and  $r_{hadr}$
are comparable values.

Below, we extend our consideration to the light meson decay constants.

\section{Light Meson Decay Constants}

The decay constant $f_P$ is defined by the matrix element:
\begin{eqnarray}
i f_P p_\mu
= \langle 0|J_\mu(0)|U(p) \rangle\,,
\label{decay0}
\end{eqnarray}
where $J_\mu$ is the axial vector current and $U(p)$ is the
normalized state vector.

There have been made many attempts to extract $f_P$ from meson
spectroscopy. Particularly, the decay constant $f_{\pi^+}$ for
$\pi^+$-meson is determined from the combined rate for
$\pi^{+}\rightarrow \mu^{+}\nu_\mu+\mu^{+}\nu_\mu\gamma$.
With the recent experimental data one obtains \cite{PDG2006}:
$$
f_\pi^{exp}=130.7\pm 0.1 \pm 0.36 {\mbox{\rm MeV}}\,, \qquad
f_K^{exp}=159.8\pm 1.4 \pm 0.44 {\mbox{\rm MeV}}\,,
\qquad f^{exp}_K/f^{exp}_\pi \simeq 1.22\pm 0.02 \,.
$$
Recent lattice QCD model with exact chiral symmetry estimates \cite{chiu05}:
$$
f^{latt}_K=152\pm 6\pm 10 {\mbox{\rm MeV}} \,.
$$
The unquenched lattice QCD calculation predicts \cite{bern05}
$$
(f_K/f_\pi)_{unquen} =1.198 \pm 0.019 \,.
$$

Following our approach, we estimate $f_P$ as follows:
\begin{eqnarray*}
i f_P~p_\mu \!\!\!&=& \!\!\! {g\over 6} \int\! {dk\over(2\pi)^4}
\int\! dx~e^{-ikx} U(x)\sqrt{D(x)}
\mbox{\rm Tr} \left\{i\gamma_5 \tilde{S}\left(\hat{k}+\mu_1\hat{p}\right) i\gamma_5
\gamma_\mu \tilde{S}\left(\hat{k}-\mu_2\hat{p}\right)\right\} \,.
\label{decay}
\end{eqnarray*}
It is reasonable to choose a normalized trial function \cite{ganb06}:
\begin{eqnarray}
\label{testf}
U(x)\sim \sqrt{D(x)}~
\exp\left\{-{a\Lambda^2 x^2/4}\right\}\,, \qquad
\int dx \left| U(x)\right|^2 =1\,, \qquad a>0\,.
\label{trial1}
\end{eqnarray}

\begin{figure}[ht]
 \centerline{
 \includegraphics[width=100mm,height=80mm]{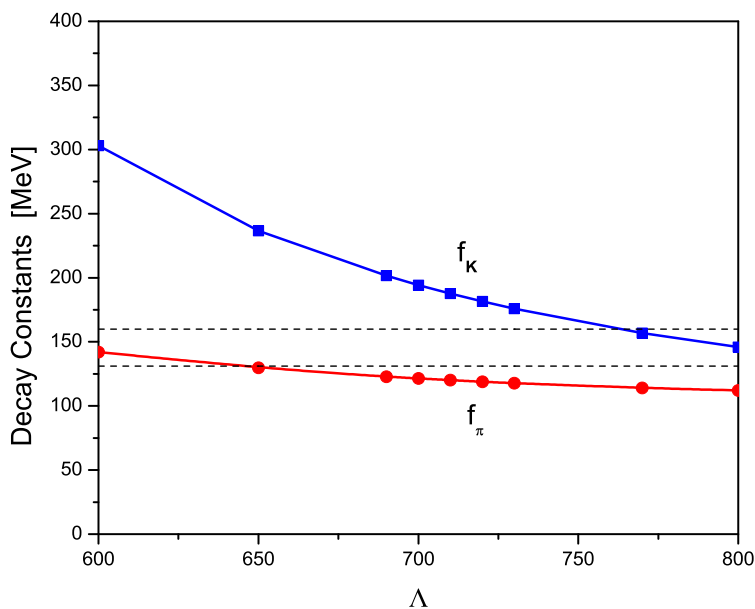} }
 \caption{Evolution of $f_\pi$ and $f_K$ with  $\Lambda$.
 Dashed lines depict the experimental data.}
\end{figure}

Omitting details of intermediate calculations write the final expression
\begin{eqnarray}
\label{decay1}
f_P \!\!\!&=& \!\!\!
{\sqrt{\alpha_s}~(m_1+m_2)\over 12\pi^{3/2} (m_1/\Lambda)^2 (m_2/\Lambda)^2}
~\exp\left\{ {m_1^2+m_2^2\over 2\Lambda^2}
\left( {M_P^2\over (m_1+m_2)^2} -1\right) \right\}  \\
&& \!\!\!\!\! \cdot \max\limits_{1/2<c<1} \left\{ (3c-1)(1-c)
~{2 m_1 m_2 + c(m_1-m_2)^2 \over (m_1+m_2)^2}
\exp\left[-{cM_P^2\over 4\Lambda^2} \left({m_1-m_2 \over m_1+m_2}\right)^2
\right] \right\} \,. \nonumber
\end{eqnarray}

The solution of Eq.(\ref{decay1}) for different values of $\Lambda$
is plotted in Fig.2 in comparison with the experimental data
\cite{PDG2006}. In the region $650\div 760 {\mbox{\rm MeV}}$ we
underestimate $f_\pi^{exp}$ and overestimate $f_K^{exp}$. Evolutions
of $f_K/f_\pi$ and  $\alpha_s$ with $\Lambda$ are depicted in Fig.3.
\begin{figure}[ht]
 \centerline{
 \includegraphics[width=80mm,height=65mm]{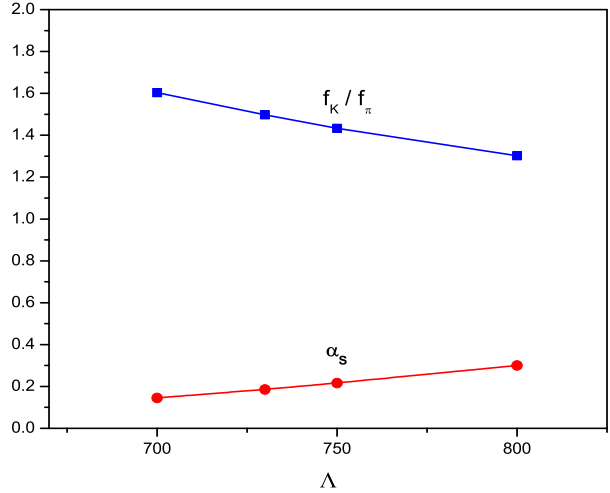} }
 \caption{The ratio $f_K/f_\pi$ and the coupling strength
 $\alpha_s$ versa the  confinement scale $\Lambda$.}
\end{figure}

For consistency, we use the parameters (\ref{parameters}) and obtain:
$$
f_\pi=118 {\mbox{\rm MeV}} \,, \qquad f_K=176 {\mbox{\rm MeV}} \,.
$$
Our estimates lie near to the latest data \cite{PDG2006,chiu05,bern05}.

\vskip 3mm

In conclusion, we have considered a relativistic quantum field model
of interacting quarks and gluons under the analytic confinement
and solve the Bethe-Salpeter equation in the one-gluon exchange
approximation for the quark-antiquark $(q\bar{q'})$ bound states.
Despite the simplicity, this model resulted in a quite reasonable
sight to the underlying physical principles of the hadronization
mechanism and gave an accurate description of the ground states,
orbital ($\ell > 0$) and radial ($n_r > 0$) excitations of mesons
in a wide range of mass (up to 9.5~GeV) and acceptable values of
the light meson decay constants $\{f_\pi,~f_K\}$. Our model provides
a solid framework to compute the meson spectra, the lowest glueball
state and decay constants of light mesons from the basic principles
of QCD and QFT.



\end{document}